\documentclass{desyproc}

\begin{document}
\title{Spectroscopic Constraints on (pseudo-)Scalar Particles  from Compact Astrophysical Objects}

\author{{\slshape Doron Chelouche$^{1,2}$, Ra\'ul Rabad\'an$^2$, Sergey S. Pavlov$^3$, \& Francisco Castej\'on$^4$}\\[1ex]
$^1$Canadian Institute for Theoretical Astrophysics,
University of Toronto, 60 Saint George Street, Toronto, ON M5S 3H8,
Canada\\
$^2$School of Natural Sciences, Institute for Advanced
Study, Einstein Drive, Princeton 08540, USA \\
$^3$Institute of Plasma Physics, Kharkov Institute of Physics and Technology, Kharkov, Ukraine \\
$^4$Laboratorio Nacional de Fusi\'on por ConÞnamiento Magn\'etico, Asociaci\'on Euratom-Ciemat para Fusi\'on, 28040 Madrid, Spain
}

\contribID{chelouche\_doron}

\desyproc{DESY-PROC-2008-02}
\acronym{Patras 2008} 
\doi  

\maketitle

\begin{abstract}

We propose a new method to search for light (pseudo-)scalar particles in the spectra of compact astrophysical objects such as magnetars, pulsars, and quasars. On accounts of compact astrophysical objects having intense magnetic fields extending over large volumes, they provide good conditions for efficient photon-particle oscillations via the Primakoff process. In particular, we show that if the coupling constant for light ($m_a<10^{-2}$\,eV) axions, $g>10^{-13}\,{\rm GeV}^{-1}$ then it is likely that absorption-like features would be detectable in the spectrum of compact astrophysical sources.

\end{abstract}

\section{Why are Compact Astrophysical Objects interesting?}

Compact astrophysical sources, by our definition, are those objects whose size is of order their 
Schwartzschild radius and include black holes and neutron stars. Neutron stars are thought to be the end stages of (not too massive) stars wherein the stellar core of the progenitor object collapsed dragging with it magnetic field lines from large volumes (and amplifying them via dynamo effects) resulting in a highly magnetized object \cite{magnet}. Another example is that of supermassive black holes which grow by matter accretion over cosmic times during which phase they become very luminous and give rise to the quasar phenomenon \cite{quasar}. While neutron star masses are of order a solar mass, supermassive black holes at the centers of galaxies may be several billion solar masses. 

The photon-axion conversion probability is given, in the limit of small values, by (e.g., \cite{osci}):
\begin{equation}
P_{\gamma \to a}\simeq \frac{1}{4}g^2B^2R^2
\label{Eq:p}
\end{equation}
where $B$ is the magnetic field intensity and $R$ is the coherence size of the field which, for compact objects, is also the size of the system. It is instructive to compare the expected $P_{\gamma \to a}$ for compact objects and for terrestrial experiments: for the CAST experiment the product, $B^2R^2$ is of order $10^{16}\,{\rm [G^2\cdot cm^2]}$.  In comparison, for pulsars whose magnetic field is of order $10^{13}$\,G and whose size is $\sim 10$\,km (e.g., \cite{pulsars}), this product is of order $\sim 10^{36}\,{\rm [G^2\cdot cm^2]}$! Therefore, pulsars are many orders of magnitude more efficient in converting photons to axions (and vice-versa).  As such, one can potentially probe down to much lower values of $g$ using compact objects than is possible using CAST (see e.g.,  \cite{raffelt} and references therein).

\section{Spectral Oscillation Features}

Efficient photon-axion oscillations occur when the momentum transfer in the conversion process is negligible. This occurs when the effective mass of the photon, as it propagates through a refractive medium (e.g., magnetized vacuum or plasma), equals that of the axion. As the refractive index for photons is frequency dependent, this condition is met only for certain frequencies and is of resonance nature. In particular, it necessitates the presence of plasma (for which the refractive index is smaller than unity) and does not occur in pure vacuum.  

For a medium with uniform conditions (e.g., plasma density, magnetic field intensity), the photon energy where resonances occur,
\begin{equation}
\omega_0=\omega_p\frac{B}{B_c} \sqrt{\frac{F_{\vert \vert}(\omega_0)-m_a^2/\omega_p^2}{7\alpha /45 \pi}}.
\label{Eq:res}
\end{equation}
$\omega_p$ is the plasma frequency, $B_c\simeq 4\times 10^{13}$\,G is the critical magnetic field, $F_{\vert \vert}$ is the normalized refractive index for photons whose polarization is parallel to the direction of the (projected) magnetic field, and $\alpha$ is the fine structure constant (this expression holds for sub-critical magnetic fields; for the more general case see \cite{chelouche}).  Resonance occurs only if $F_{\vert \vert}-m_a^2/\omega_p^2 > 0$ which sets an upper limit on the axion mass which can be efficiently probed by a given environment. For $m_a/\omega_p \ll 1$ and setting $F_{\vert \vert}=1$ (as appropriate for cold plasma), we find that resonances are expected to occur around infrared to optical energies for pulsars, hard X-ray and gamma ray energies for quasars, and sub-mm to infrared energies for magnetars (note that the above expression does not hold for magnetars for which $B>B_c$; see \cite{chelouche} for the general case).

The photosphere of compact objects emits over a broad spectral range and one therefore expects to detect photon deficits at photon-axion resonance energies. Phenomenologically, such features would look like absorption features although the physics is very different.  Making accurate spectral predictions for such features requires one to follow the evolution of the coupled photon-axion system  with time from the photosphere where photons are created to the telescope. The details of the equations of motion involved and their solution are fully discussed in \cite{chelouche}. We only mention that the important refractive processes to include are due to plasma (whether active or inactive, cold or hot) as well as vacuum birefringence.

Here we focus on one example being that of magnetars.  In those objects, the magnetic field and plasma density, $\rho$, vary with distance from the photosphere such that $B\propto \rho \propto R^{-3}$ (dipolar field configuration; \cite{pulsars}).  Judging from Eq. \ref{Eq:res}, it is clear that photons of different energy will undergo resonance conversion at different locations in the atmosphere. In particular, the predicted spectral feature is broad as $B$ and $\rho$ decay monotonically with distance with $\omega_0$ spanning a wide energy range.  For all cases considered here the variations in the magnetospheric conditions are slow enough (adiabatic) so that considerable photon to axion conversion occurs (see \cite{dong}). 

\begin{figure}[]
\centerline{\includegraphics[width=0.51\textwidth]{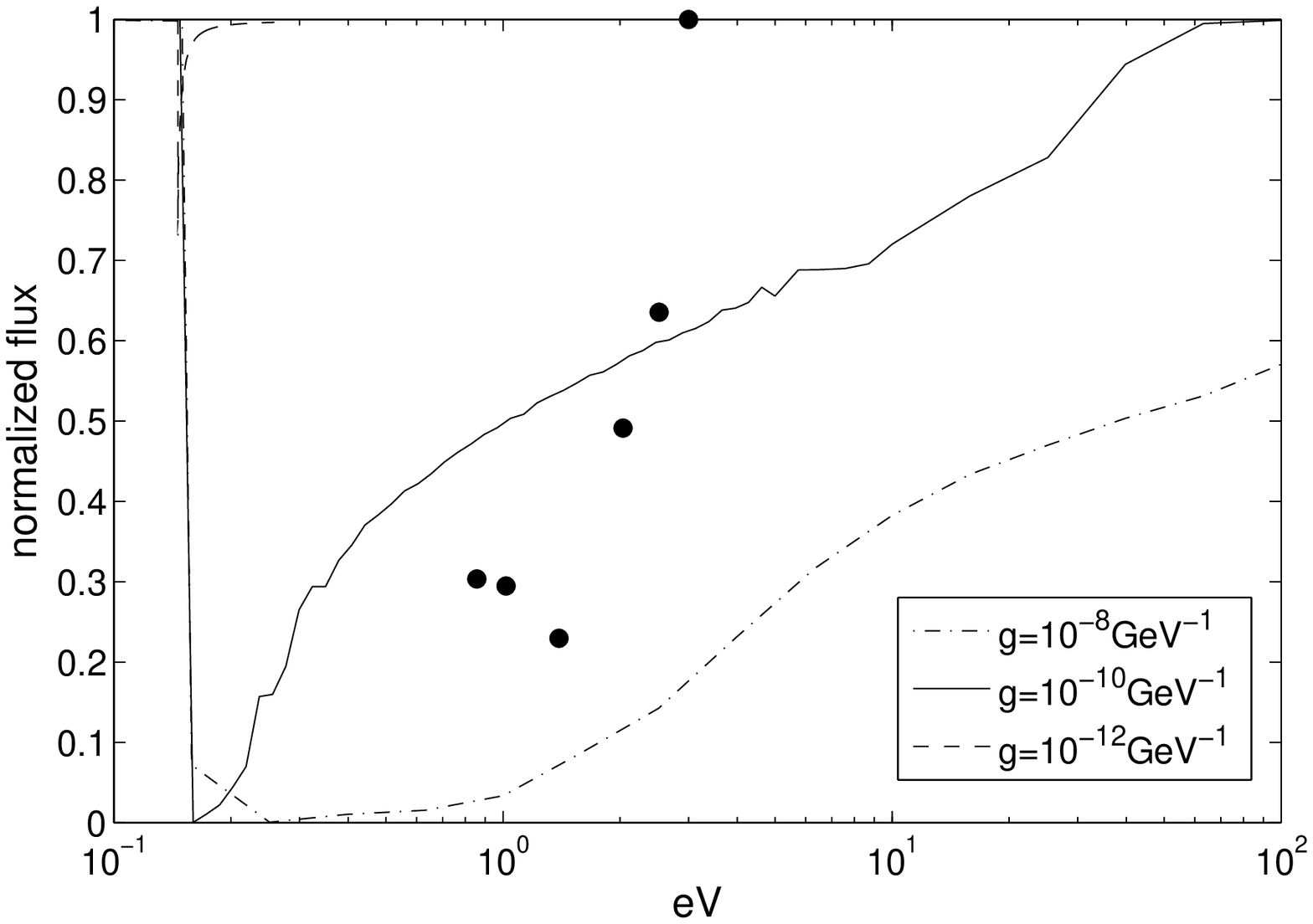}\includegraphics[width=0.5\textwidth]{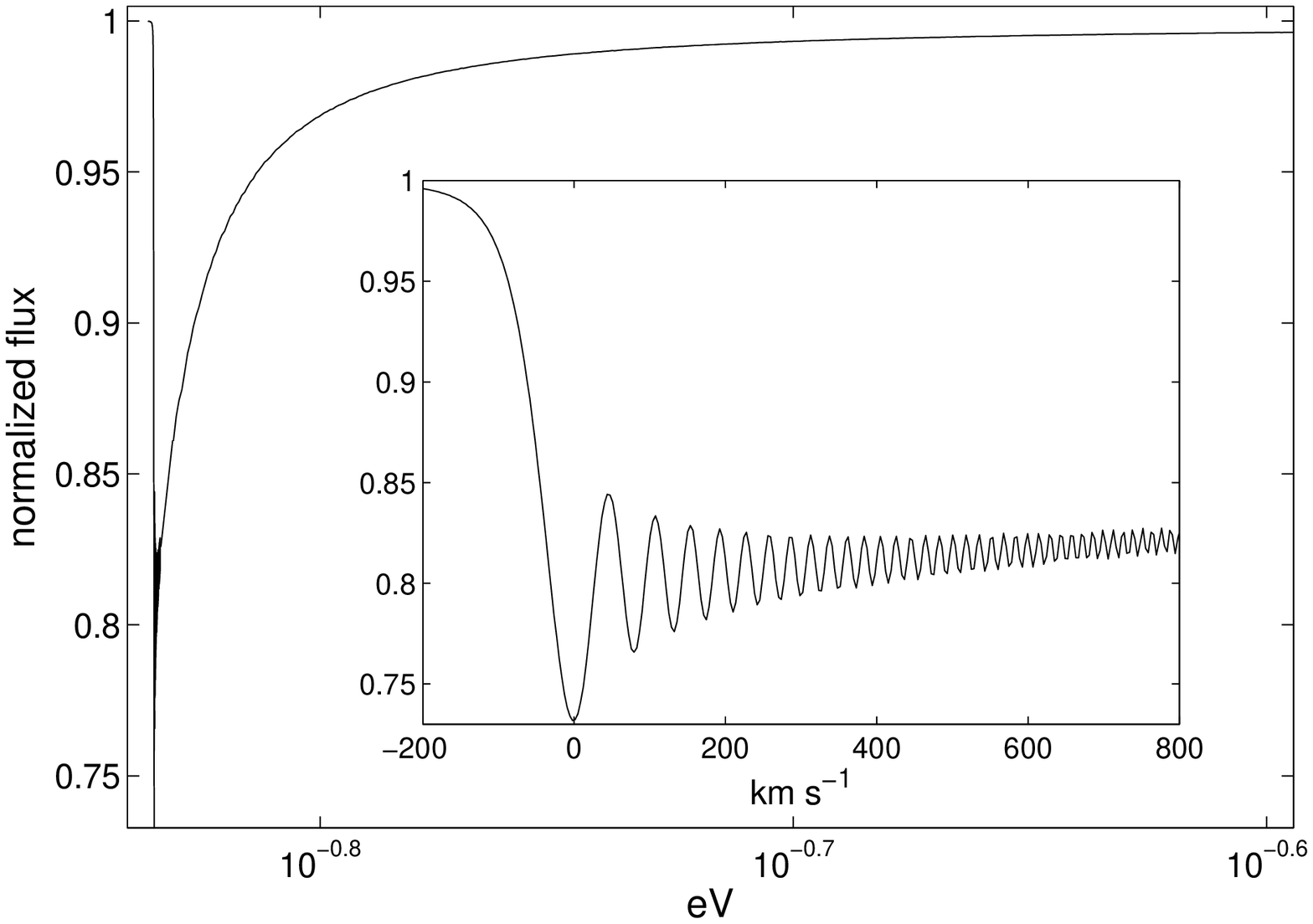}}
\caption{{\it Left:} Spectral oscillation feature as predicted for magnetars ($B=10^{15}$\,G and $n(r_\star)=10^{15}\,{\rm cm^{-3}}$ (with the photospheric radius, $r_\star=10$\,km) for several values of $g$. Note the prominent absorption-like feature which can be observed for $g=10^{-10}\,{\rm GeV}^{-1}$ which is the best current limit on $g$ from CAST and horizontal branch stars. Also shown for comparison is (normalized) photometric data for a magnetar (circles; adopted from \cite{magnetar}). {\it Right:} Zoom in on the spectral oscillation feature for $g=10^{-12}\,{\rm GeV}^{-1}$. The feature is broad and can be detected using low resolution spectra. The inset shows a high resolution blow up of the feature revealing oscillatory pattern being a trademark of photon-particle spectral conversion features. }
\label{Fig:spectrum}
\end{figure}

Some examples of the predicted spectral oscillation features are shown in Fig. 1 for several values of $g$. At the current upper limit given by CAST of $g\sim 10^{-10}\,{\rm GeV}^{-1}$, a very broad absorption-like spectral feature should be  easily detectable (extending from infrared to optical energies in the example shown).  For lower values of $g$ ($\sim 10^{-12}\,{\rm GeV}^{-1}$), the feature is narrower yet can be easily detected using  broad band photometry and low to medium resolution spectra. Using high resolution spectra one may be able to identify the oscillatory pattern, a trademark of photon-particle conversion and is not expected to arise from atomic emission and absorption processes (see Fig. 1). In cases where the properties of the object along the line of sight change with time (e.g., a rotating star), the spectral oscillation feature would be time-dependent. In particular, the energy and strength of the spectral feature would vary with rotation phase of the star and phase-sensitive observations may be crucial for correctly identifying it. In addition, plasma composition, photon polarization, magnetic field direction with respect to the photon propagation direction as well as the plasma temperature could all  affect the predicted spectral oscillation features. A full treatment of these issues is given in \cite{chelouche}.

In the above, we have given spectral predictions for the specific case of magnetars. However, a similar line of reasoning applies also to pulsars, magnetars, and other types of compact objects. In all cases relatively broad spectral oscillation features are predicted and can be detected down to low values of $g$. Such spectral features may be identified by {\bf (a)} their broad band shape, {\bf (b)} the presence of small energy-scale oscillations, {\bf (c)} distinct variability pattern which is intimately connected with time variations in the mean magnetic field strength and plasma density in the magnetosphere along the line of sight.  In particular, the spectral properties of such features are very different from those characterizing atomic features.

\section{Implications for Axion Physics}

\begin{figure}[]
\centerline{\includegraphics[width=0.55\textwidth]{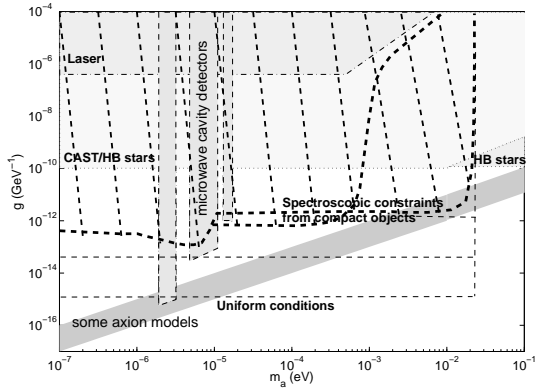}}
\caption{The particle parameter space which can be probed by observations of compact astrophysical sources assuming flux decrements of 10\% can be detected. Clearly, the proposed spectroscopic method can probe down to much lower values of $g$ (for axion mass of $< 10^{-2}$\,eV it is possible to probe down to $g\lesssim 10^{-12}\,{\rm GeV}^{-1}$) than is accessible by other methods (thick hatched dashed region). Also, the proposed method does not require axions to be dark matter particle. For comparison we show the parameter space probed by systems of size $r_\star$ with uniform magnetic and plasma density size (thin hatched dashed region). See \cite{chelouche}  for further details.}
\label{Fig:param}
\end{figure}

The calculated spectral oscillation features for compact objects depend on their intrinsic properties. Assuming that the current understanding of such systems is qualitatively correct then it is possible to define the axion parameter space which can be probed by spectroscopic studies of these sources (more details can be found in \cite{chelouche}). Assuming one can detect flux decrements of order 10\% in the data (as noted before, this requires low to medium resolution spectra), the sensitivity limit in $g$ and $m_a$ is shown in Fig. 2. Clearly, this method can, in principal, increase our sensitivity to light ($m_a<10^{-2}$\,eV) axions by 2-3 orders of magnitude using certain types of objects. Furthermore, the regions of parameter space which can be probed by different classes of objects overlap allowing to reduce systematic effects and corroborate the results. We emphasize that the proposed method works equally well for scalar and pseudo-scalar particles and does not require axions to be dark matter candidates.



\begin{footnotesize}



%

\end{footnotesize}


\end{document}